\documentclass[conference]{IEEEtran}
\IEEEoverridecommandlockouts

\usepackage{cite}
\usepackage{amsmath,amssymb,amsfonts}
\allowdisplaybreaks
\usepackage{algorithmic}
\usepackage{graphicx}
\usepackage{caption}
\usepackage{subcaption}
\usepackage{textcomp}
\usepackage{xcolor}
\usepackage{comment}
\begin{document}

\title{Energy Consumption of Electric Vehicles: Effect of Lateral Dynamics}


\author{Simran Kumari\textsuperscript{1}\thanks{This work is supported by Ministry of Education (MoE), Govt. of India under Prime Minister Research Fellowship (PMRF) scheme and partially supported by the project HEV of SRIC IIT Kharagpur, jointly funded by Tata Motors Limited (TML) and Govt. of India under UAY scheme.}\thanks{Authors are with Department of Electrical Engineering, Indian Institute of Technology (IIT) Kharagpur, India. Email: \textsuperscript{1}simranjnr@kgpian.iitkgp.ac.in, \textsuperscript{2}susenjit@iitkgp.ac.in, \textsuperscript{3}ahota@ee.iitkgp.ac.in and \textsuperscript{4}smukh@ee.iitkgp.ac.in.} \and Susenjit Ghosh\textsuperscript{2} \and Ashish R. Hota\textsuperscript{3} \and Siddhartha Mukhopadhyay\textsuperscript{4}}

\maketitle

\begin{abstract}
Current research on energy related problems such as eco-routing, eco-driving and range prediction for electric vehicles (EVs) primarily considers the effect of longitudinal dynamics on EV energy consumption. However, real-world driving includes longitudinal as well as lateral motion. Therefore, it is important to understand the effects of lateral dynamics on battery energy consumption. This paper conducts an analysis of the stated effect and validates its significance through simulations. Specifically, this study demonstrates that inclusion of the effect of lateral dynamics can improve accuracy and reliability of solutions in eco-routing, eco-driving and range prediction applications. 
\end{abstract}


\section{Introduction}
EV technology has seen a boom in recent years \cite{kumar2017development}. However, state-of-art technology does not adequately address the issue of range anxiety among EV drivers. Various reasons leading to this issue are route, traffic, driver and vehicle powertrain \cite{varga2019prediction}. There have been several works related to EV energy consumption such as energy consumption prediction \cite{morlock2019forecasts}, \cite{yi2017adaptive}, \cite{desreveaux2019impact}, eco-driving \cite{lee2020model},\cite{shen2020minimum},\cite{dib2014optimal}, eco-routing \cite{yi2018optimal},\cite{chakraborty2021intelligent} and range prediction \cite{varga2019prediction},\cite{oliva2013model}, \cite{ondruska2014probabilistic} in order to deal with range anxiety issue. EV energy consumption model is a fundamental block for these applications. 

State-of-art works use different approaches for developing the same. Earlier work \cite{fiori2016power} presents quasi-steady backward power-based energy consumption model and computes regenerative braking efficiency. Additionally, \cite{ristiana2019new} presents a dynamic model namely integrated battery-electric vehicle model which include battery dynamics, motor dynamics as well as vehicle longitudinal dynamics. Reference \cite{han2019fundamentals} presents analysis of energy optimal driving for conventional as well as electric vehicles from optimal control perspective. It utilizes Pontryagin’s Minimum Principle (PMP) to obtain velocity profile which minimizes wheel to distance and tank to distance energy consumption. In \cite{wang2015optimal}, authors study motion control problems such as cruise distance maximization and travel time minimization utilizing electric vehicle power consumption model for an EV. Additionally, \cite{dib2014optimal} models eco-driving problem as battery energy consumption minimization problem over road segments. Furthermore given predicted drive cycle and current state of charge (SOC), \cite{oliva2013model} utilizes unscented kalman filter (UKF) to predict SOC profile through quasi-static power consumption model. Above works  utilize longitudinal vehicle dynamics for modelling EV energy consumption. However, longitudinal vehicle dynamics does not capture realistic on-road driving scenario due to various factors such as road geometry, driver intention and traffic behaviour. Realistic driving includes coupled longitudinal and lateral motion. Therefore, it is necessary to consider the effect of steering along with accelerator and brake pedal actuators to obtain more accurate estimate of energy consumption.

Few recent literature have attempted to capture the effect of steering action on energy consumption. For given turning radius, \cite{ding2018optimal} estimates lateral force for different values of longitudinal velocity and finds the energy optimal velocity for achieving turning maneuver. Based on terminal optimal velocity values, transition velocity profile between straight and curved road is calculated through a dynamic programming approach. In \cite{li2019energy}, authors evaluate maximum cornering speed at which tire force does not saturate and  use this in hybrid electric vehicle (HEV) energy management. However, these works do not incorporate varying curvature road and lane changes which are common in real-life driving situations. In order to address this research gap, we present an analysis of the effect of lateral dynamics on energy consumption of a rear wheel driven (RWD) EV. Our analysis indicates that state of art energy consumption models based on longitudinal motion underestimate energy consumption in EVs, and motivates inclusion of lateral dynamics in such models.

\section{Energy Flow in an Electric Vehicle}
It is important to understand the flow of energy from energy source to maneuver in order to analyse the effect of lateral dynamics on energy consumption. An EV powertrain generally consists of battery as energy storage, motor as propulsion source followed by fixed gear differential with its axles attached to wheel. The flow of energy for a RWD EV maneuvering from time $t_0$ to $t_f$ is presented below.

\begin{figure}
\centerline{\includegraphics[trim={0 0 0 1cm},clip,width=0.5\textwidth,height=5cm,keepaspectratio]{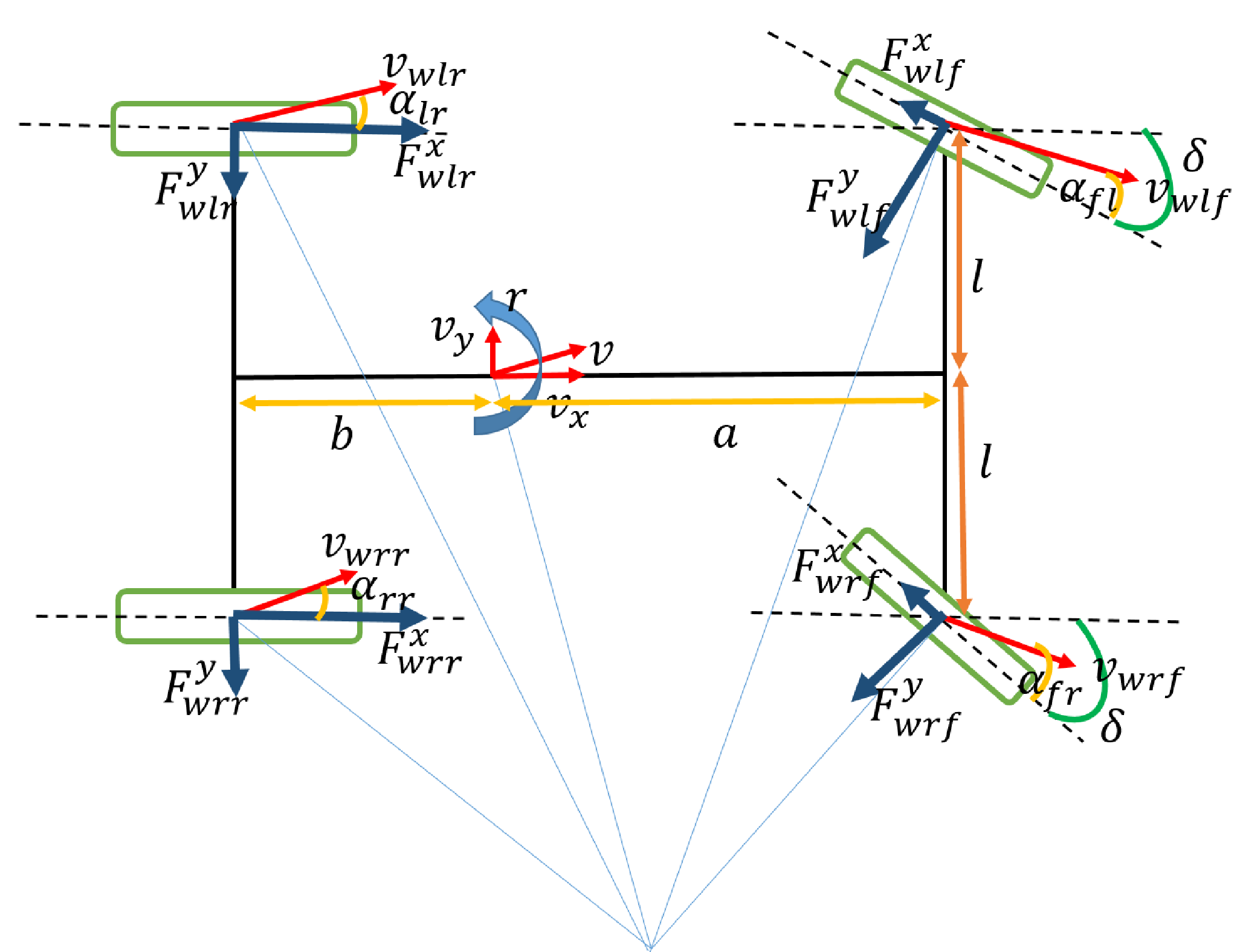}}
\caption{{\footnotesize Schematic of different quantities associated with dual track bicycle model.}}
\label{dyn}
\end{figure}

\subsection{Energy consumed in maneuver} 
Different quantities associated to EV dynamics during maneuver are shown in Fig.~\ref{dyn}. The distance of center of gravity (COG) from front and rear axles are denoted by $a$ and $b$ respectively. Additionally, $l$ denotes half of track width of the EV and EV front wheels are steered by an angle $\delta$. Longitudinal velocity, lateral velocity and yaw rate of COG of the EV are denoted by $v_x$, $v_y$ and $\Dot{\psi}$ respectively. Dependence on time is suppressed for better readability. Similarly, $F_x$, $F_y$ and $M_z$ denote resultant longitudinal force, resultant lateral force at and resultant yaw moment about COG of the EV. The force and moment are generated by forces acting at wheel-road contact. When an EV is performing coupled longitudinal and lateral maneuver, utilized energy is given as: 
\begin{align*}
E_{w,m}&=\int_{t_0}^{t_f}F_xv_x\,dt+\int_{t_0}^{t_f}F_yv_y\,dt+\int_{t_0}^{t_f}M_z\Dot{\psi}\,dt.
\end{align*}
Here, $E_{w,m}$ is energy utilized in maneuver. Assuming that camber angle of each wheel is zero, the relationship between forces and moment at vehicle level with wheel forces are: 
\begin{align*}
    F_x=&(F_{wlr}^x+F_{wrr}^x)-(F_{wlf}^y+F_{wrf}^y)\sin{\delta}\\
    &-(F_{wlf}^x+F_{wrf}^x)\cos{\delta}-F_a,\\
        F_y=&(F_{wlf}^y+F_{wrf}^y)\cos{\delta}+(F_{wlr}^y+F_{wrr}^y)\\
        &-(F_{wlf}^x+F_{wrf}^x)\sin{\delta},\\
        M_z=&(-F_{wlr}^y-F_{wrr}^y)b+(F_{wlf}^yl\sin{\delta}+F_{wlf}^ya\cos{\delta})\\
        &+(F_{wrf}^yl\sin{\delta}+F_{wrf}^ya\cos{\delta})\\
        &+(F_{wrf}^xl\cos{\delta}-F_{wrf}^ya\sin{\delta})\\
        &+(F_{wlf}^xl\cos{\delta}-F_{wlf}^ya\sin{\delta}). \label{eq1}
\end{align*}
Here, $F_{w*\#}^x$ and $F_{w*\#}^y$ are wheel longitudinal and lateral forces at $*\#$ wheel where $*\in \{l:\text{left},r:\text{right}\}$ and $\#\in \{f:\text{front}, r:\text{rear}\}$. Throughout the paper, subscripts $*\in \{l:\text{left},r:\text{right}\}$ and $\#\in \{f:\text{front}, r:\text{rear}\}$ are used. Similar to forces, wheel velocities are also related to vehicle velocities and the relationship is given as:
\begin{align*}
        v_{wlf}^x&=(v_x-\Dot{\psi}l)\cos{\delta}+(v_y+\Dot{\psi}a)\sin{\delta},\\ v_{wlf}^y&=-(v_x-\Dot{\psi}l)\sin{\delta}+(v_y+\Dot{\psi}a)\cos{\delta},\\
        v_{wrf}^x&=(v_x+\Dot{\psi}l)\cos{\delta}+(v_y+\Dot{\psi}a)\sin{\delta},\\
        v_{wrf}^y&=-(v_x+\Dot{\psi}l)\sin{\delta}+(v_y+\Dot{\psi}a)\cos{\delta},\\
        v_{wlr}^x&=v_x-\Dot{\psi}l, v_{wlr}^y=v_y-\Dot{\psi}b,\\
        v_{wrr}^x&=v_x+\Dot{\psi}l, v_{wrr}^y=v_y-\Dot{\psi}b,
  \end{align*}
  where, $v_{w*\#}^x$ and $v_{w*\#}^y$ are longitudinal and lateral components of wheel velocity at wheel $*\#$.

\subsection{Energy consumed in wheel translation motion} Brake torque $T_{b*\#}$ and rolling resistance moment $T_{R*\#}$ resist wheel rotational motion. However, axle torque $T_{w*\#}$ assist in wheel rotation during acceleration and resist the rotational motion during deceleration in presence of regenerative braking. Therefore, input energy consumed by wheel is 
    \begin{align*}
    E_{w*\#,in}&=\int_{t_0}^{t_f}(T_{w*\#}-T_{b*\#}-T_{R*\#})\omega_{w*\#}\,dt.
    \end{align*} 
    Some part of wheel input energy is used in rotating wheel $E_{rot,w*\#}$ and wheel longitudinal maneuver $E_{w*\#,out}^x$  while rest is lost due to friction $E_{loss,w*\#}$. Energy consumed in wheel longitudinal translation motion is given as:
\begin{align*}
  E_{w*\#,out}^x&=E_{w*\#,in}-E_{rot,w*\#}-E_{loss,w*\#}\\&=\int_{t_0}^{t_f}(F_{w*\#}^xv_{w*\#}^x)dt,\text{where,}\\
        E_{rot,w*\#}&=\int_{t_0}^{t_f}(T_{w*\#}-T_{b*\#}-T_{R*\#}-F_{w*\#}^xr_w)\omega_{w*\#}\,dt\\
        E_{loss,w*\#}&=\int_{t_0}^{t_f}F_{w*\#}^x(r_w\omega_{w*\#}-v_{w*\#}^x)\,dt.       
\end{align*}
Here, $\omega_{w*\#}$ and $r_w$ denote wheel rotational speed and radius of wheel (same for all the wheels).
 Wheel lateral forces, generated at tyre-road contact due to steering action, resist wheel motion in order to align with wheels in steered direction and thus dissipate energy. Therefore, lateral forces also contribute to wheel output energy along with longitudinal forces. The output energy of individual wheel is:
\begin{align*}
E_{w*\#,out}=&E_{w*\#,out}^{x}+E_{w*\#,out}^{y}\\
=&\int_{t_0}^{t_f}(F_{w*\#}^xv_{w*\#}^x+F_{w*\#}^yv_{w*\#}^y)\,dt
        \end{align*}
        where, $E_{w*\#,out}^{y}$ is energy consumed in wheel lateral translation motion. For RWD vehicles, $E_{w*\#,out}^x$ is positive for rear wheels which contribute to wheel traction energy. The value of $E_{w*\#,out}^x$ is negative for front wheels indicating dissipation of energy due to resisting wheel longitudinal forces during acceleration as well as deceleration. This causes decrease in traction energy and leads to a lower value of resultant traction energy. In absence of regeneration $E_{rot,w*\#}$ and $E_{w*\#,out}^x$ are lost as thermal energy during braking. However in presence of regenerative braking, energy is not totally lost as thermal energy and some part is recovered as battery energy.
        
Through simplification of equations, it can be shown that energy consumed in maneuver is same as sum of energy consumed in translation motion of all wheels. Thus,
\begin{align*}
E_{w,m}&=E_{wlr,out}+E_{wrr,out}+E_{wlf,out}+E_{wrf,out}.
\end{align*}

\subsection{Input energy from powertrain to wheel} Axle torque contributes to input energy of wheel from powertrain $E_{w*\#,in}^p$ and is given as:
    \begin{align*}
    &E_{w*\#,in}^p=\int_{t_0}^{t_f}(T_{w*\#}\omega_{w*\#})\,dt,
      \end{align*}
    where $T_{wlf}=T_{wrf}=0$. Input energy for individual wheels are given below:
    \begin{align*}
         &E_{w*\#,in}=E_{w*\#,in}^p+\int_{t_0}^{t_f} (-T_{b*\#}-T_{R*\#})\omega_{w*\#}\,dt.
    \end{align*}
    For RWD EV, differential output energy, denoted by $E_{d,out}$, is distributed into rear left and right wheels as: 
    \begin{align*}
           E_{d,out}&=E_{wlr,in}^p+E_{wrr,in}^p.
    \end{align*}
    
    \subsection{Energy flow from battery to differential}
    Motor output energy is input to differential and is related to differential output energy as $E_{d,out}=\eta_{diff}E_{m,out}$. Here, $\eta_{diff}$ and $E_{m,out}$ are differential efficiency and motor output energy respectively. Motor output energy is given as:
    \begin{align*}
      E_{m,out}&=\int_{t_0}^{t_f} T_m(t)\omega_m(t)\,dt\\
      &=E_{m,in}-\int_{t_0}^{t_f}P_{m,loss}(T_m(t),\omega_m(t))\,dt,\\
      \text{where} \quad E_{m,in}&= \begin{cases}
            \eta_{inv}E_{b,out}, \quad T_m\geq 0\\
            \frac{E_{b,out}}{\eta_{inv}}, \quad T_m \leq 0.
        \end{cases}
    \end{align*}
    Here, $T_m$, $\omega_m$, $\eta_{inv}$, $P_{m,loss}$, $E_{m,in}$ and $E_{b,out}$ are motor torque, motor speed, efficiency of inverter, motor power loss, motor input energy and battery output energy respectively.
    
    With $V_b$ and $I_b$ as terminal voltage and current flowing through battery, output energy of battery is:
    \begin{align*}
        E_{b,out}= \int_{t_0}^{t_f} V_b(t)I_b(t) \,dt.
    \end{align*}
It is evident from above energy flow analysis that energy from battery is utilized in rotating wheel and to overcome friction loss, front wheels slippage loss (due to component of resistive wheel longitudinal forces) and cornering loss (due to component of resistive wheel lateral forces). Therefore, overall energy demand for an EV trying to achieve specific longitudinal maneuver over a duration increases in presence of lateral maneuver. Similar analysis can be applied to front wheel driven (FWD) EV.


\subsection{Energy Flow Simulation Results}
\begin{figure}[htbp]
    \centering
    \includegraphics[trim={0 0 0 0cm},clip,width=0.45\textwidth,keepaspectratio]{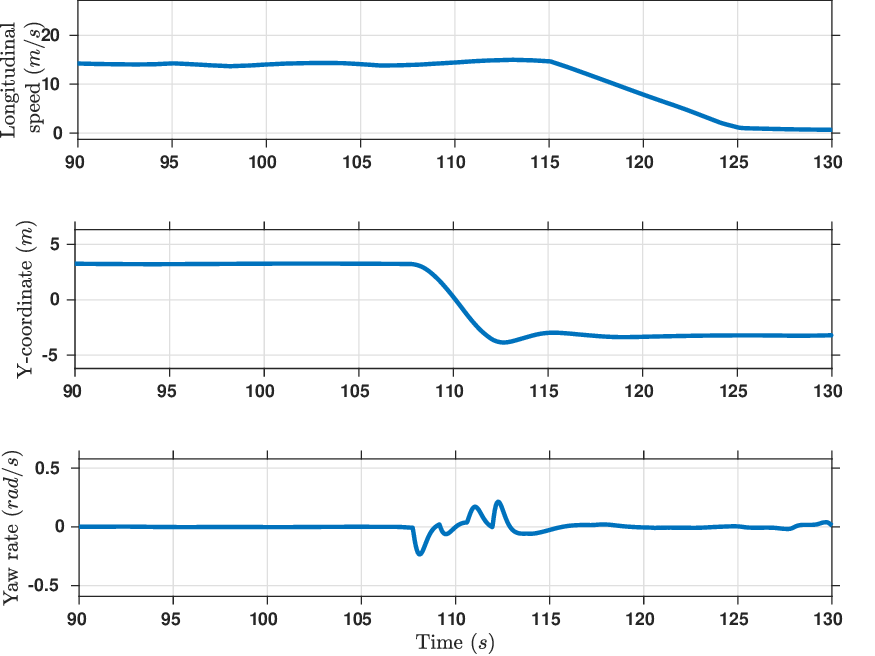}
    \caption{{\footnotesize Maneuver of EV}}
    \label{fig:maneuver}
\end{figure}

\begin{figure}[htbp]
\centering
\begin{subfigure}{0.45\textwidth}
\includegraphics[width=\textwidth,keepaspectratio]{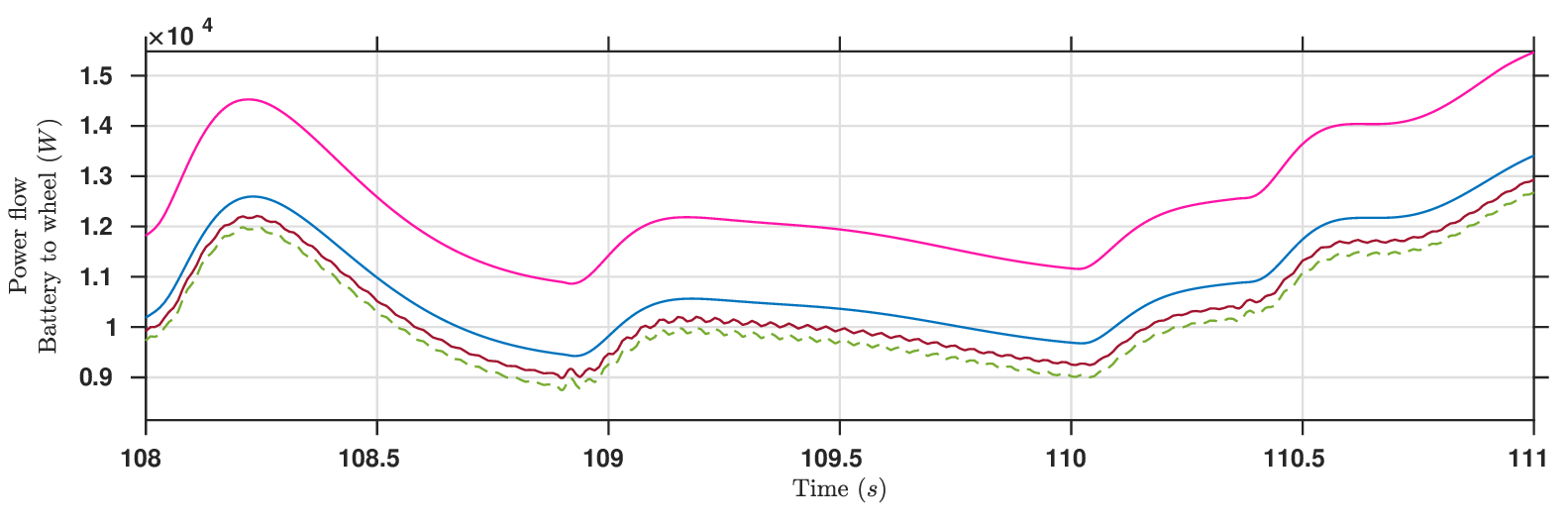}
\caption{{\footnotesize Power flow from battery to wheel. \textcolor{magenta}{Magenta}: battery output power, \textcolor{blue}{blue}: motor output power, \textcolor{purple}{purple}: differential output power and \textcolor{green}{green}: Input power to wheels from powertrain.}}
\end{subfigure}
\hfill
\begin{subfigure}{0.45\textwidth}
\includegraphics[width=\linewidth,keepaspectratio]{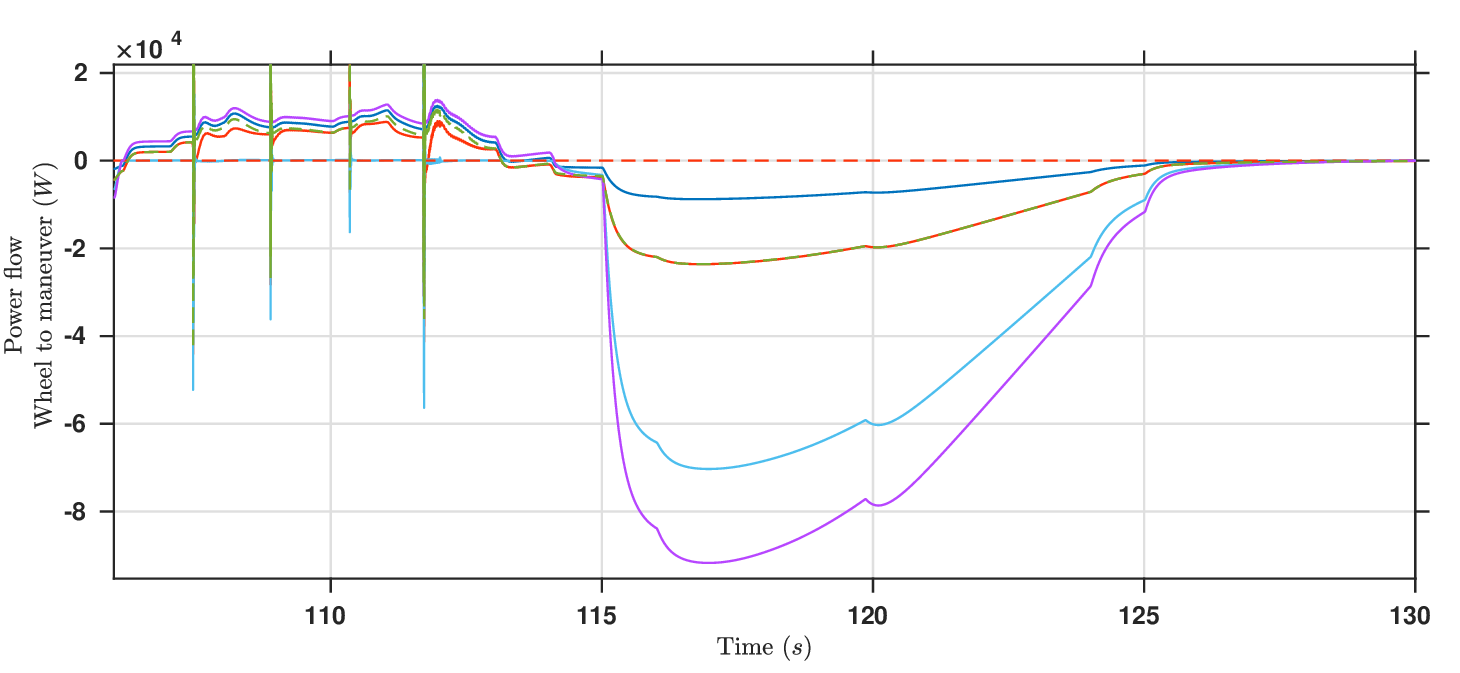}
\caption{{\footnotesize Power flow from wheel to maneuver. \textcolor{violet}{Violet}: Input power to wheels, \textcolor{cyan}{cyan}: Power corresponding to energy consumed in rotating wheels, \textcolor{blue}{blue}: rear wheels traction power, \textcolor{green}{green}: resultant wheel traction power, \textcolor{orange}{orange}: Power corresponding to friction loss at wheels and \textcolor{red}{red}: power consumed in achieving maneuver.}}
\end{subfigure}
\caption{{\footnotesize Power flow in absence of regeneration for maneuver shown in Fig.~\ref{fig:maneuver}.}}
\label{lat}
\end{figure}

\begin{figure}[htbp]
\centering
\begin{subfigure}{0.45\textwidth}
\includegraphics[width=\textwidth,keepaspectratio]{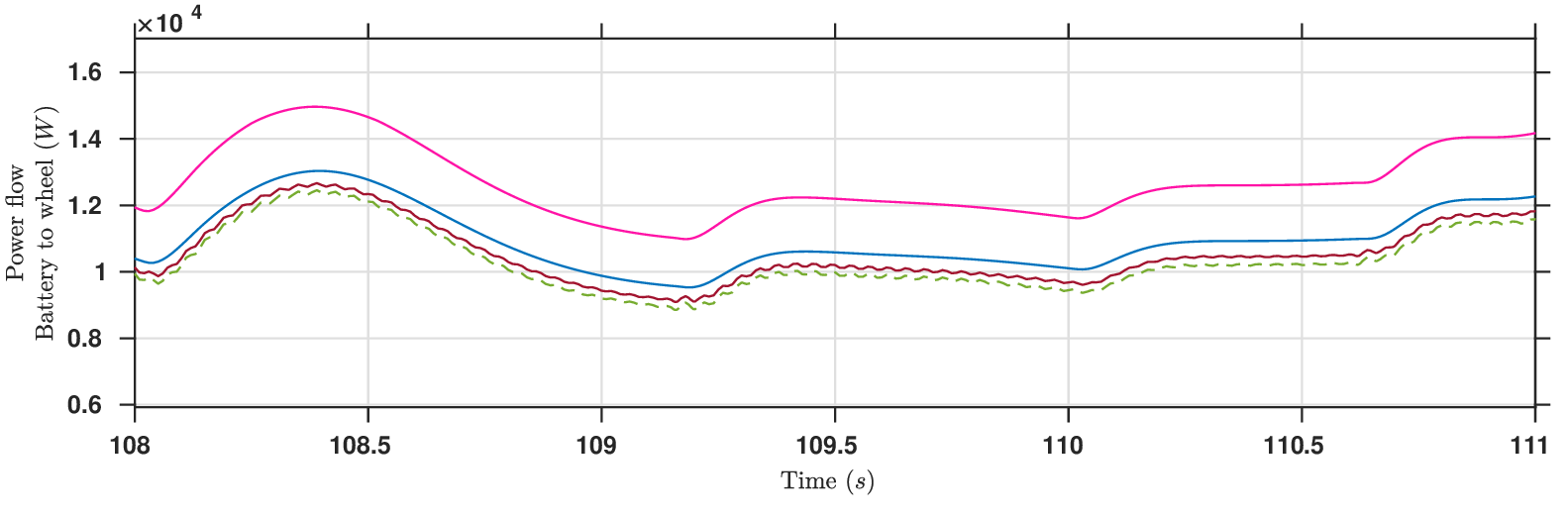}
\caption{{\footnotesize Power flow from battery to wheel. \textcolor{magenta}{Magenta}: battery output power, \textcolor{blue}{blue}: motor output power, \textcolor{purple}{purple}: differential output power and \textcolor{green}{green}: Input power to wheels from powertrain.}}
\end{subfigure}
\hfill
\begin{subfigure}{0.45\textwidth}
\includegraphics[width=\textwidth,keepaspectratio]{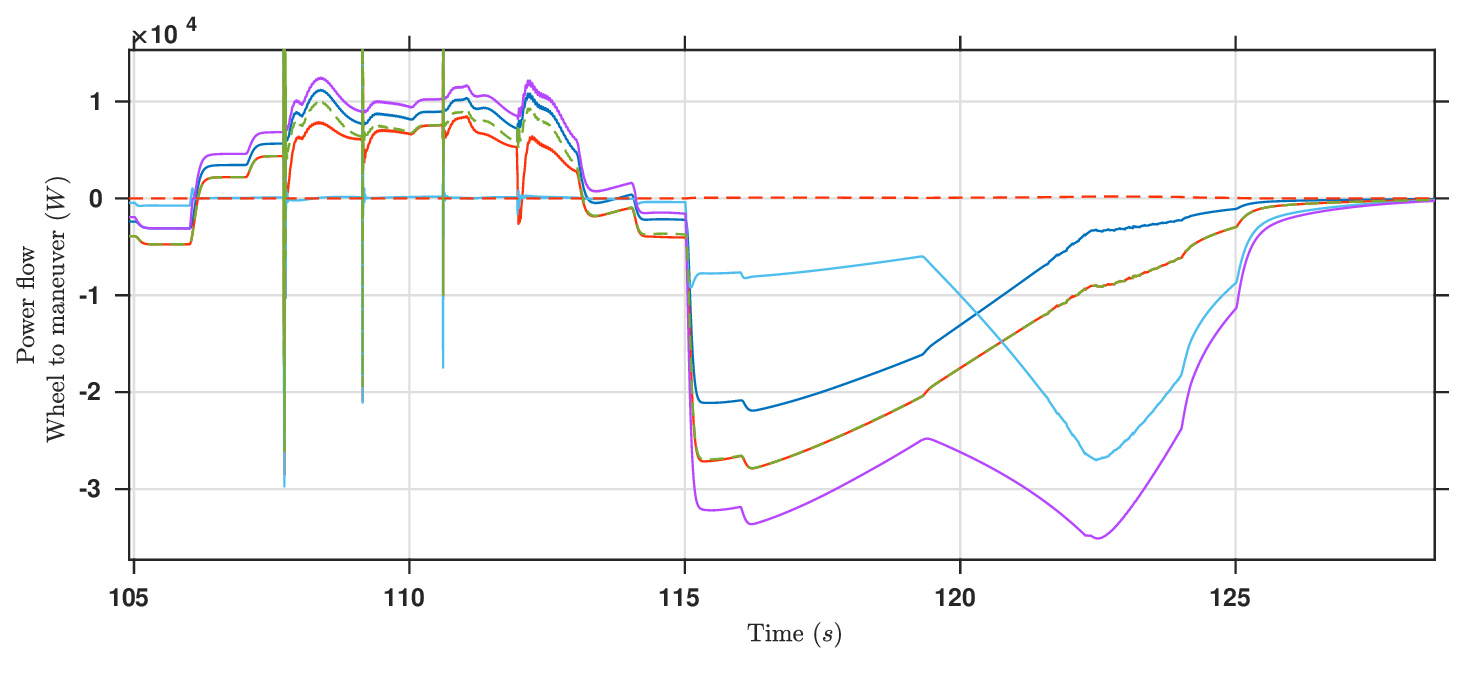}
\caption{{\footnotesize Power flow from wheel to maneuver. \textcolor{violet}{Violet}: Input power to wheels, \textcolor{cyan}{cyan}: Power corresponding to energy consumed in rotating wheels, \textcolor{blue}{blue}: rear wheels traction power, \textcolor{green}{green}: resultant wheel traction power, \textcolor{orange}{orange}: Power corresponding to friction loss at wheels and \textcolor{red}{red}: power consumed in achieving maneuver.}}
\end{subfigure}
\caption{{\footnotesize Power flow in presence of regeneration for maneuver shown in Fig.~\ref{fig:maneuver}.}}
\label{lat1}
\end{figure}

The presented flow of energy can be visualized graphically in terms of power flow from source to maneuver.\footnote{Analysis upto next section is carried out for a specific driver behaviour. Results for various driver behavior would qualitatively remain the same.} To analyze the energy flow, a simulation study is conducted using Matlab-Simulink \cite{MATLAB:2021}. An EV model, consisting of powertrain and planar dynamics, is simulated to track FTP-75 drive cycle for a given driver behaviour. The planar maneuver followed by EV is characterized by longitudinal speed, inertial Y-coordinate and yaw rate. Fig.~\ref{fig:maneuver} shows snapshot of maneuver for duration $90-130s$.
Power flow profile for this small duration of the EV maneuver in absence and presence of regeneration braking is shown in Fig.~\ref{lat} and \ref{lat1} respectively. 
Positive battery power indicates situation when driver is pressing accelerator pedal and energy flows from battery to maneuver. In this case, power corresponding to input energy consumed by wheels is smaller in magnitude compared to battery power due to efficiency factor of motor and differential. Resisting front wheel longitudinal forces causes decrease in magnitude of wheel traction power and leads to resultant traction power with smaller magnitude. The power consumed in maneuver is same as resultant traction power in case of longitudinal motion. However, there is a decrease in power consumed during lateral maneuver due to resistive front wheel lateral forces. Power consumed in wheel rotation and dissipated in tyre-road contact friction loss is at most 2\% of power supplied by battery and is negligible. However, the magnitude of  power consumed in wheel rotation is not negligible during braking. Instead negative value of this power and power consumed in maneuver, in Fig.~\ref{lat}, indicate that brake mechanism dissipates vehicle maneuver energy and wheel rotational energy as thermal energy. In case of regeneration, negative wheel input power and battery power, shown in Fig.~\ref{lat1}, indicates some part of wheel rotation and maneuver energy flows back to battery and rest is dissipated as thermal energy during braking.

\begin{figure}[htbp]
\begin{subfigure}{\textwidth}
\includegraphics[width=0.45\textwidth,keepaspectratio]{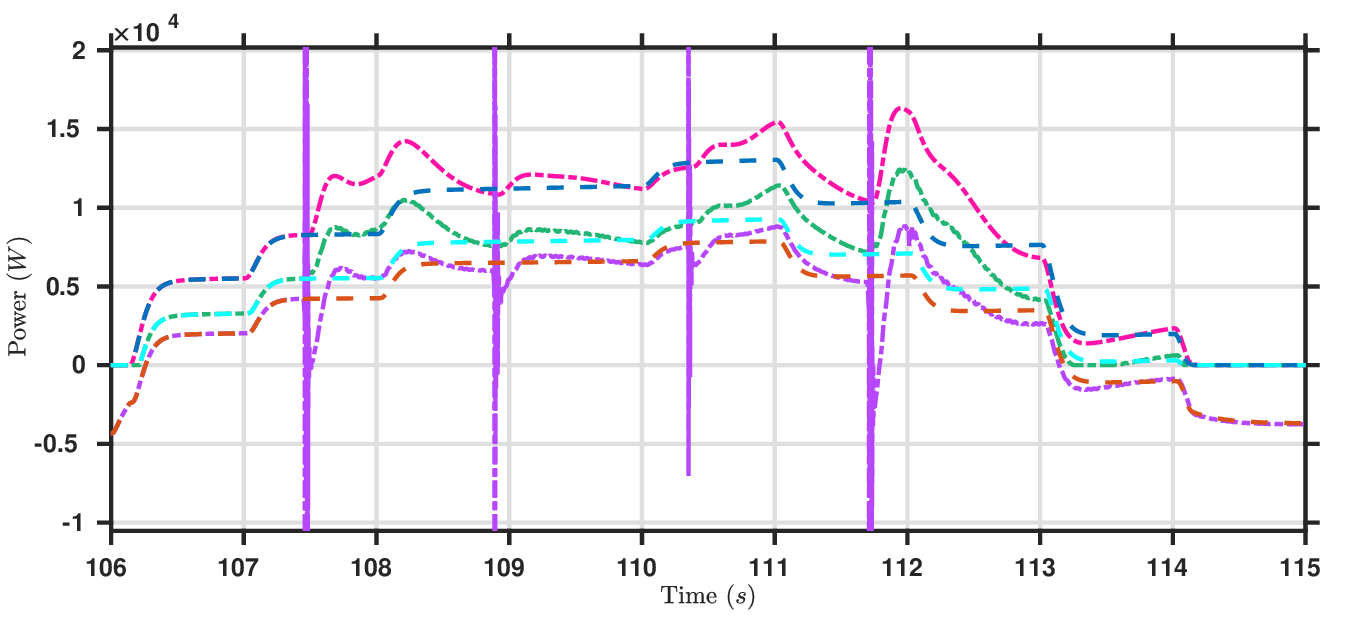}
\end{subfigure}
\hfill
\begin{subfigure}{\textwidth}
\includegraphics[width=0.45\textwidth,keepaspectratio]{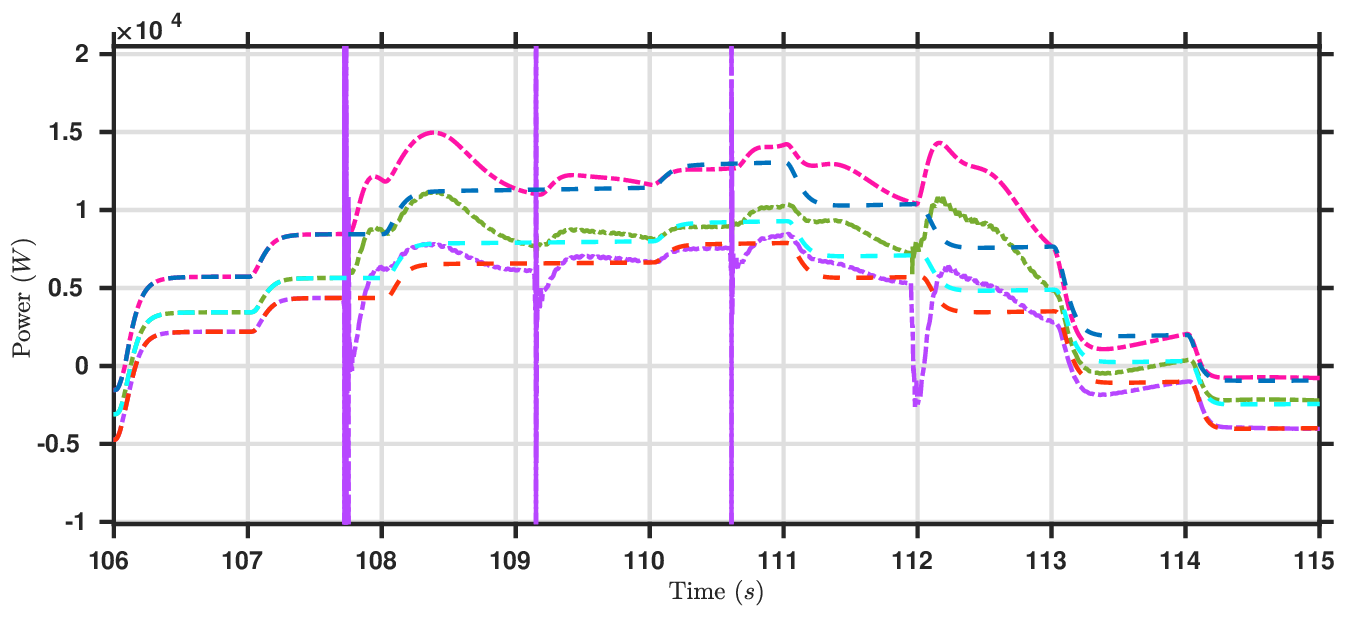}
\end{subfigure}
\caption{{\footnotesize Power profile variation for EV maneuver with and without lane change in absence (top) and presence (bottom) of regeneration. Dashed: absence of lane change and dash-dotted: presence of lane change. \textcolor{magenta}{Magenta}, \textcolor{blue}{blue}: battery output power, \textcolor{green}{green}, \textcolor{cyan}{cyan}: traction power and \textcolor{violet}{violet}, \textcolor{orange}{orange}: power consumed in maneuver.}}
\label{energydiff1}
\end{figure}

Fig. \ref{energydiff1} shows the variation in power requirement of an EV trying to achieve the longitudinal maneuver specified in Fig~\ref{lat} in absence and presence of specified lane change maneuver. The plot in the top panel corresponds to EV without regeneration and the plot in the bottom panel corresponds to EV with regeneration. It can be observed that power demand increases in presence of lateral maneuver. There is a decrease in longitudinal speed during lateral maneuver due to resisting wheel lateral forces. Driver presses accelerator pedal more to sustain target speed and battery power demand increases as a result. It can also be observed that power demand is less in case of regeneration. It is evident from the discussion that EV lateral dynamics has an impact on battery energy consumption. Therefore, the next section analyzes significance of this impact on energy consumption of an EV driving over long range.


\section{Effect of Lateral Dynamics on EV Range}
\begin{figure}[htbp]
    \centering
    \includegraphics[trim={0cm 0 0 0cm},clip,width=0.5\textwidth,keepaspectratio]{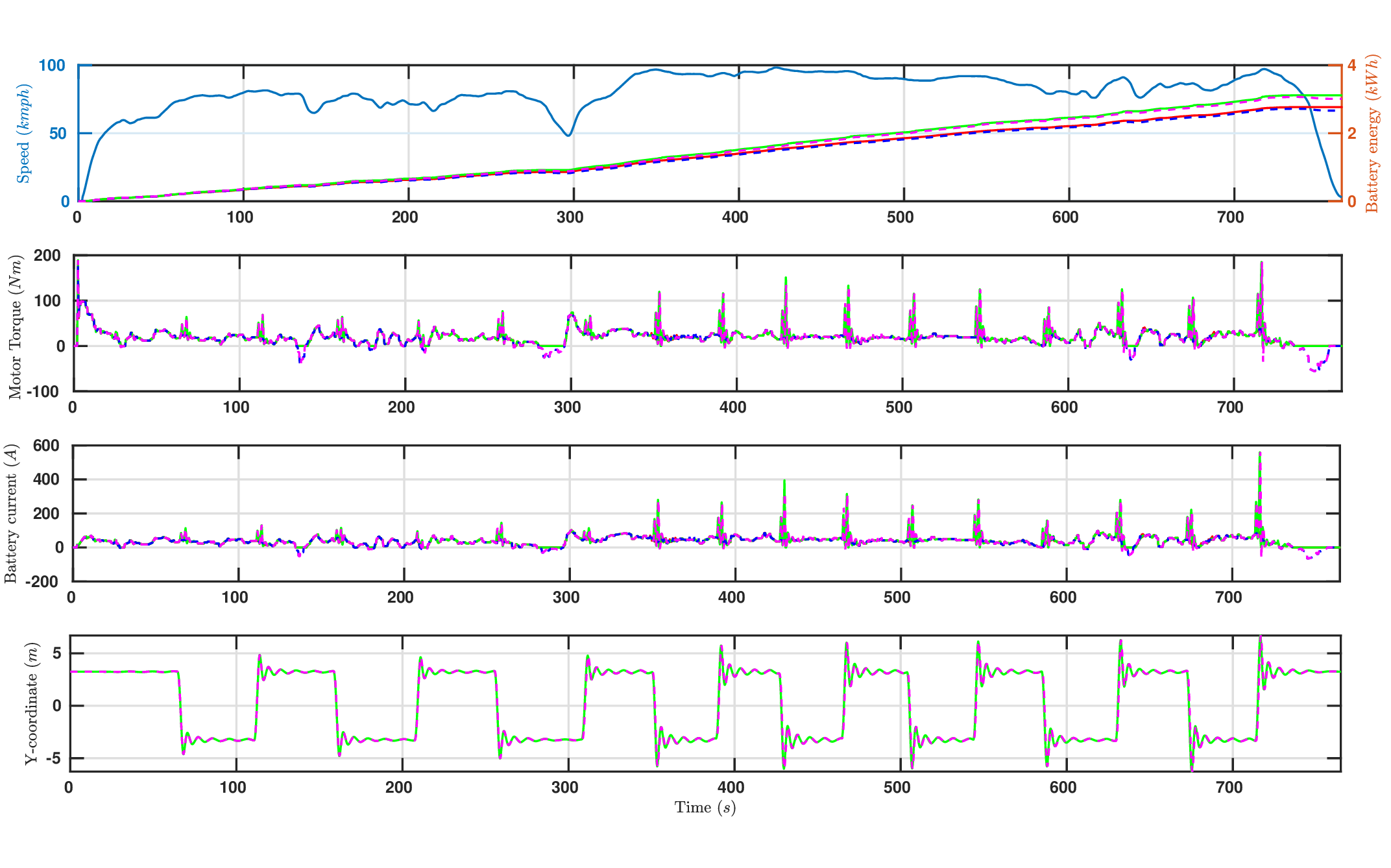}
    \caption{\footnotesize Energy profile for HWFET drive cycle. Here, \textcolor{red}{Red}: Without lane change and regeneration, \textcolor{green}{green}: With lane change and without regeneration, \textcolor{blue}{blue}: Without lane change and with regeneration and \textcolor{magenta}{magenta}: With lane change and regeneration.}
    \label{energy}
\end{figure}

To analyze significance of the effect of lateral dynamics on energy consumption, an EV model with given driver behavior is simulated to track standard drive cycles in two different cases. The two cases are: first is longitudinal maneuver and second is longitudinal maneuver with a lane change every 1km approximately. It can be observed from Fig.~\ref{energy} that energy consumed for HWFET drive cycle in latter case is more compared to former for EV without regeneration barking. Corresponding motor torque profile has large peak at every lane change indicating increase in energy demand for EV maneuver. Regeneration braking provides EV with facility of energy recuperation. Therefore, energy consumed in presence of regeneration barking is less compared to one without it. The opportunity of recuperation for EV in latter case is less compared to former. Since, cornering forces already assist in braking, amount of negative torque required for braking is less. Similar observation is obtained for other drive cycles such as NEDC and FTP 75.
Various parameters relating energy consumption corresponding to different drive cycles to EV performance are included in Table \ref{tab1}-\ref{tab3}. Driving range of EV is obtained through extrapolation of consumed battery energy over repetition of drive cycle for a $54.28$ kWh battery. It can be observed that there is a significant decrease in driving range when lane changes are included in maneuver. Thus, neglecting lateral dynamics gives an overestimated range. 
\begin{table}[htbp]
\caption{Various Parameters for NEDC}
\begin{tabular}{|p{1.75cm}|p{1.25cm}|p{1.25cm}|p{1.25cm}|p{1.25cm}|}
\hline
&\multicolumn{4}{|c|}{\textbf{NEDC}}\\
\cline{2-5} 
\textbf{Parameters}&\multicolumn{2}{|c|}{\textbf{Without Regeneration}} &\multicolumn{2}{|c|}{\textbf{With Regeneration}}\\
\cline{2-5}
 & \textbf{\textit{Without lane changes}}& \textbf{\textit{With lane changes}}& \textbf{\textit{Without lane changes}}&\textbf{\textit{With lane changes}} \\
\hline
Distance($km$)& $11.764$ & $11.768$ & $11.715$ & $11.718$\\ \hline
Battery Energy($kWh$)& $1.9$ & $2.035$ & $1.7$ & $1.823$\\ \hline
Range($km$) & $336.08$ & $313.82$ & $373.86$ & $348.94$ \\ \hline
\end{tabular}
\label{tab1}
\end{table}

\begin{table}[htbp]
\caption{Various Parameters for FTP 75}
\begin{tabular}{|p{1.75cm}|p{1.25cm}|p{1.25cm}|p{1.25cm}|p{1.25cm}|}
\hline
&\multicolumn{4}{|c|}{\textbf{FTP 75}}\\
\cline{2-5} 
\textbf{Parameters}&\multicolumn{2}{|c|}{\textbf{Without Regeneration}} &\multicolumn{2}{|c|}{\textbf{With Regeneration}}\\
\cline{2-5}
 & \textbf{\textit{Without lane changes}}& \textbf{\textit{With lane changes}}& \textbf{\textit{Without lane changes}}&\textbf{\textit{With lane changes}} \\
\hline
Distance($km$)& $19.35$ & $19.264$ & $19.216$ & $19.15$\\ \hline
Battery Energy($kWh$)& $3.101$ & $3.23$ & $2.652$ & $2.708$\\ \hline
Range($km$) & $338.64$ & $323.63$ & $393.32$ & $383.8$ \\ \hline
\end{tabular}
\label{tab2}
\end{table}

\begin{table}[htbp]
\caption{Various Parameters for HWFET}
\begin{tabular}{|p{1.75cm}|p{1.25cm}|p{1.25cm}|p{1.25cm}|p{1.25cm}|}
\hline
&\multicolumn{4}{|c|}{\textbf{HWFET}}\\
\cline{2-5} 
\textbf{Parameters}&\multicolumn{2}{|c|}{\textbf{Without Regeneration}} &\multicolumn{2}{|c|}{\textbf{With Regeneration}}\\
\cline{2-5}
 & \textbf{\textit{Without lane changes}}& \textbf{\textit{With lane changes}}& \textbf{\textit{Without lane changes}}&\textbf{\textit{With lane changes}} \\
\hline
Distance($km$)& $16.93$ & $16.94$ & $16.9$ & $16.92$\\ \hline
Battery Energy($kWh$)& $2.77$ & $3.114$ & $2.66$ & $3.01$\\ \hline
Range($km$) & $331.92$ & $295.16$ & $344.45$ & $304.92$ \\ \hline
\end{tabular}
\label{tab3}
\end{table}
\section{Conclusion}
An analysis of energy flow in an EV is carried out and it is observed that there is an increase in demanded energy to achieve specific longitudinal profile in presence of lateral maneuver. It is revealed that state-of-art energy consumption model underestimates the energy consumption profile of an EV during maneuver and lateral dynamics has significant impact on EV performance in real-world driving situation. Therefore, it is essential to include its effect in energy consumption model for different applications such as eco-driving, range prediction, etc. Further analysis to incorporate effect of various environmental factors such as road friction, slope and wind speed during planar maneuver of EV remains a promising direction for future research.
\bibliographystyle{ieeetr}
\bibliography{paper_4_final_submission}
\end{document}